# Evaluating current processors performance and machines stability

R. Esposito, P. Mastroserio, G. Tortone
*INFN, Napoli, I-80126, Italy*

F. M. Taurino
*INFM, Napoli, I-80126, Italy*

Accurately estimate performance of currently available processors is becoming a key activity, particularly in HENP environment, where high computing power is crucial. This document describes the methods and programs, opensource or freeware, used to benchmark processors, memory and disk subsystems and network connection architectures. These tools are also useful to stress test new machines, before their acquisition or before their introduction in a production environment, where high uptimes are requested.

## 1. INTRODUCTION

The "benchmarking suite" shown in this paper consists in some free applications used to evaluate and stress test each machine subsystem: CPU, memory hierarchy, disks, network.

## 2. BENCHMARKING SUITE

### 2.1. CPU and memory

**GLIBENCH**

This tool executes Dhrystones (MIPS), Whetstones (MFLOPS), Matrix operations, Number crunching, Floating point and Memory throughput tests [1].

**NBENCH**

Based on beta release 2 of BYTE Magazine's BYTEmark benchmark program (previously known as BYTE's Native Mode Benchmarks), and runs 10 tests to compare the running machine with an AMD K6 @233MHz. It returns a three indexes: Memory, Integer and Floating-point [2].

**BYTEBENCH**

Used to test a *nix machine in different ways. It runs arithmetic tests, system tests like process spawning or context switching [3].

**LMBENCH**

It's a series of micro benchmarks intended to measure basic operating system and hardware system metrics. The benchmarks fall into three general classes: bandwidth, latency, and "other" [4].

**UBENCH**

Ubench is executing rather senseless mathematical integer and floating-point calculations for 3 mins concurrently using several processes, and the result is Ubench CPU benchmark. It is executing rather senseless memory allocation and memory to memory copying operations for another 3 mins concurrently using several processes, and the result is Ubench MEM benchmark [5].

**MEMPERF**

It measures the memory bandwidth in a 2 dimensional way. First it varies the block size which provides information of the throughput in different memory system hierarchys (different cache levels). Secondly it varies the access pattern from contiguous blocks to different strided accesses [6].

**STREAM**

The STREAM benchmark is a simple synthetic benchmark program that measures sustainable memory bandwidth (in MB/s) and the corresponding computation rate for simple vector kernels [7].

**LLCBENCH**

It groups three benchmarks: BlasBench, to test BLAS routines; CacheBench, to test cache memory; MPBench, to test MPI implementations [8].

**APFLOAT**

It is a high performance arbitrary precision package that can be used to perform calculations involving millions of digits, such as $\pi$ [9].

**POVRAY**

This well known program creates 3dimensional graphics, using standard, Athlon optimized or Pentium optimized binaries [10].

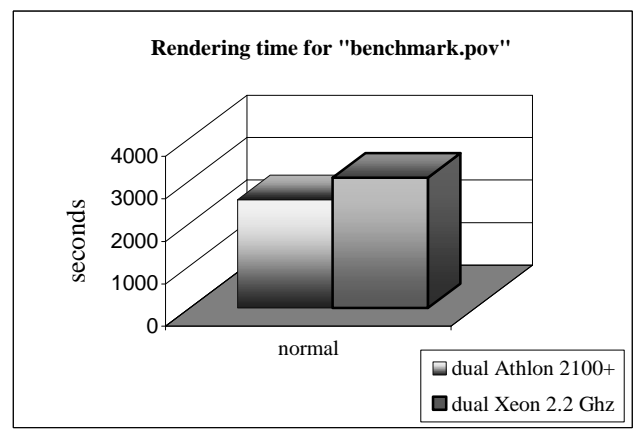

Figure 1: Example of PovRay benchmark results

## 2.2. Disks

**BONNIE++**

A modified version of Bonnie, which creates, reads, writes and deletes very big files [11].

**IOZONE**

This benchmark generates and measures a variety of file operations: read, write, re-read, re-write, read backwards, read strided, fread, fwrite, random read/write, pread/pwrite variants, aio_read, aio_write, mmap [12].

## 2.3. Network

**NETPERF**

It provides tests for both unidirecitonal throughput, and end-to-end latency with TCP, UDP, sockets [13].

**NETPIPE**

This tool can benchmark network communications with non standard hardware, like high speed interconnections used in cluster environments [14].

**PALLAS**

It's a complex benchmarks used to evaluate MPI performance. It provides a concise set of benchmarks targeted at measuring the most important MPI functions [15].

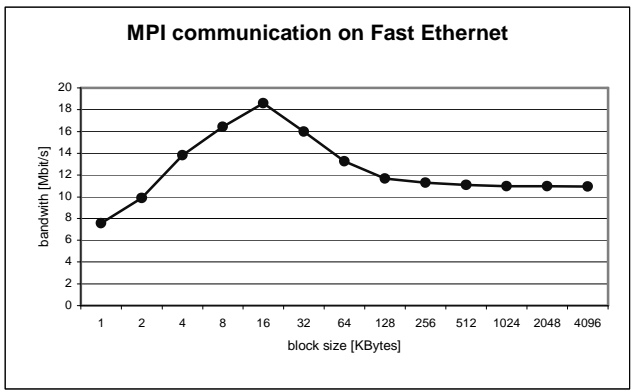

Figure 2: Example of PALLAS benchmark results

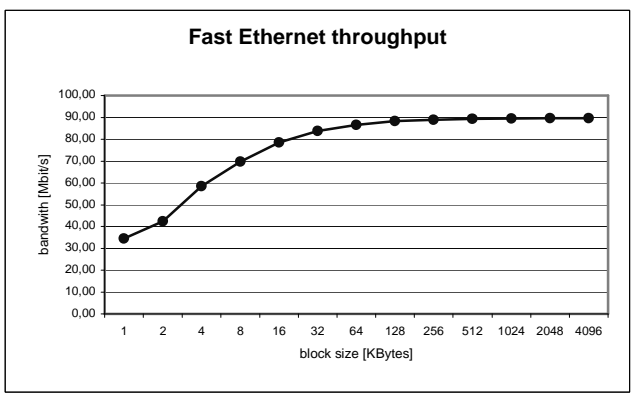

Figure 3: Example of NETPIPE benchmark results

**TUDP003**

## 3. CONCLUSIONS

This set of benchmarks allows us to accurately characterize raw performances of available machines.

Though many commercial or free benchmark tools are currently available, we have chosen the ones shown in this document because, in our experience, they seem to give a satisfying performance analysis of every single hardware component. Furthermore this suite of benchmarks has proven to be a valid tool to stress test machines before starting production activities.